\documentclass[a4paper,12pt]{article}
\usepackage[cp1251]{inputenc}
\usepackage[english,russian]{babel}

\usepackage{graphicx}
\usepackage{caption}
\DeclareCaptionLabelSeparator{fill}{{. }}
\usepackage{cite}
\usepackage{geometry}
\usepackage{hhline}
\usepackage{amsmath}
\usepackage{indentfirst} % Красная строка
\usepackage{eufrak}
\title{Энтропия дислокаций: Зависимость от температуры и плотности}
\author{	{А.Г. Сухарев } \\
\normalsize{
\begin{tabular}{c}
АО ``ГНЦ РФ Троицкий институт инновационных и термоядерных \\
 исследований'', 108840 Троицк, Москва, Россия\\
e-mail: sure@triniti.ru
\end{tabular}}
  }
\date{}

\begin{document}
	\maketitle

%	\begin{multicols}{1}
	\begin{abstract}
Лазерное упрочнение металлов происходит под воздействием ударной волны, которая меняет распределение и плотность одномерных дефектов - дислокаций. Существует связь между плотностью дислокаций, размером зерна и устойчивостью монокристалла к сдвиговой нагрузке. Механизм процессов упрочнения продолжает интенсивно изучаться, и динамика дефектов играет здесь центральную роль. В данной работе выполнен анализ энтропии дислокаций с комбинаторной точки зрения и с точки зрения физического осциллятора с заданным запасом энергии. Оба вклада играют важную роль при описании свободной энергии одномерного ансамбля дислокаций, и необходимы для учета динамических процессов внутри зерна поликристаллической структуры.           \\
{\bf{Ключевые слова:}} {\footnotesize{ одномерные дефекты, дислокации, энтропия, лазерный наклеп}}
    \end{abstract}
			
			\section{Введение}
Лазерное упрочнение металлов и сплавов инициированием ударной волны (Laser Shock Peening), получившее жаргонное название ``лазерный наклёп'', собирает заслуженный интерес из-за технологической возможности повысить усталостную прочность титановых сплавов \cite{z1,z2,z3}. Описанию технологии LSP мы посвятили достаточно много места в предыдущей работе \cite{r1}, сейчас центром нашего внимания будут термодинамические характеристики дислокаций. Учёные - материаловеды долгое время утверждали, что статистическая термодинамика неприменима к дислокациям, однако теперь точка зрения изменилась \cite{t1}. Ключевые процессы, происходящие в поликристаллической структуре металла под действием ударной волны, связаны с наличием одномерных дефектов - дислокаций. Когда они перемещаются по кристаллическому материалу, система подвергается необратимой деформации сдвига. Способность дислокаций принимать на себя нагрузку обеспечивает прочность отдельных зёрен поликристаллической структуры. Каркас дислокаций воспринимает нагрузку как единое целое, дрейф дислокаций в поле соседей и внешней нагрузки направлен в сторону конфигурации с большей вероятностью, в которой уменьшаются напряжения внутри зерна. Кроме того, кинетика дислокации происходит в поле периодического потенциала Пайерлса-Набарро, который не только локализует дислокацию в потенциальной яме, но и определяет термодинамические функции дислокаций. Термодинамические свойства решётки твёрдых тел выводятся в результате расчёта свободной энергии нормальных колебаний совокупности ряда независимых осцилляторов. Поскольку дислокации подвержены сдвигу (то есть потенциальная яма имеет конечную глубину), то их термодинамические величины, связанные с колебательной степенью свободы, следует изучать в рамках теории физического, а не математического осциллятора. Учёт влияния энергии ударной волны приводит к выводу, что именно одномерные дефекты воспринимают ударную нагрузку на себя. В результате, температура дислокаций и самой кристаллической решётки могут различаться. Поэтому для описания системы удобно разбить ее на два слабо связанных множества, одно состоит из дислокаций, второе - из колебательных мод решётки прочей части системы. С этой точки зрения пластическую деформацию описывают как движение дислокаций по зерну. А события на поверхности зерна приводят к динамике размеров зёрен. Более того процесс измельчения зерна приводит к заметному упрочнению структуры решётки \cite{t2}. Следуя \cite{t1} повторяем, что основная задача - решение проблемы упрочнения, для которой актуален вопрос, почему напряжение, необходимое для деформирования, увеличивается по мере деформирования материала. Для получения понимания процессов упрочнения на уровне физики процессов необходимо определить энтропию дислокаций. В целях упрощения рассмотрим случай одномерной структуры одинаково ориентированных в поперечном направлении дефектов. Введение в решётку дислокации увеличивает число возможных способов размещения атомов в ней \cite{t3}. Это обуславливает появление конфигурационной энтропии и изменение колебательной энтропии, так вблизи от дислокации изменяется частота колебаний атомов. В противовес точечным дефектам, у которых существует равновесная плотность для данной температуры, дислокации являются термодинамически неравновесными дефектами, т.е. их количество не фиксируется температурой. Это означает неприменимость большой статистической суммы Гиббса с переменным числом частиц \cite{t4}. Поэтому далее расчёт термодинамических величин дан при заданном числе дислокаций. В линейной цепочке связанных полем напряжений дислокаций существуют два масштаба времён, связанных с процессами нормализации температуры соседних осцилляторов и нормализации позиций дислокаций. Согласно \cite{t5}, если процесс тепло-переноса может протекать в локально-равновесных условиях и соответствует классическому случаю теплопроводности, то процесс переноса дислокаций может быть локально-неравновесен (и не подчиняется классическому уравнению диффузии). Принцип локального термодинамического равновесия означает, что несмотря на градиенты температуры, в каждом малом элементе среды существует состояние локального равновесия, для которого локальная энтропия является той же функцией макроскопических переменных, что и для равновесных систем. Состояние неравновесной системы описывается набором локальных независимых термодинамических параметров, причём есть только неявная зависимость от координаты и времени через них. Второй независимый принцип, описываемый в \cite{t5}, связан с локальностью, когда форма уравнения не зависит от характерных масштабов системы. Локальность под вопросом, когда, например, предел уменьшения объёма области ограничивается физическим размером ячейки (атома). Или, когда встаёт вопрос о эквивалентности суммы и интеграла, или производной - со своим сеточным аналогом.

Конфигурационную энергию ${U_0}(\rho )$  и энтропию ${S_0}(\rho )$  дислокаций следует относить к единице длины. Энергия ${U_0}(\rho )$ пропорциональна произведению плотности $\rho$ размещения дислокаций и энергии погонной длины дислокации ${e_d}$. Энтропия ${S_0}(\rho )$ вычисляется путём подсчёта числа возможных расположений дислокаций при фиксированных значениях ${U_0}(\rho )$ и $\rho $ \cite{t6}. Свободная энергия ${F_0} = {U_0} - {T_{eff}}{S_0}$ может быть минимизирована по отношению $\rho $. Минимум ${F_0}$ соответствует устойчивому распределению Больцмана-Гиббса с эффективной температурой ${T_{eff}}$. Барьеры для дислокаций на границе зёрен делают более вероятными состояния ближе к границе, изменяя как распределение дислокаций, так и исходную интегральную энтропию дислокаций. В работе \cite{t7} оценена энтропия пары дислокаций на кристалл из 75 и 192 атома от расстояния между ними по изменению спектра колебаний ($s = \ln \frac{{eT}}{{\hbar \bar \omega }}$) при переходе от идеального кристалла к напряжённому после добавления пары дислокаций ($\Delta s/k = 0.3$).  Причём считается, что установилось температурное равновесие между кристаллом и дислокациями. При быстром деформировании внутренняя энергия поступательно-колебательных мод дислокаций не равна температуре зерна, и цитированная модель теряет применимость.
Рассматривая подвижность дислокаций по кристаллу, мы понимаем, что она обусловлена их внутренней энергией и отношением внутренней энергии к величине барьера Пайерлса-Набарро. Для простоты представим барьер Пайерлса-Набарро в форме гармонического потенциала. В рамках такой модели у подвижности дислокаций будет фазовый переход при приближении их энергии к значению на сепаратриссе. Только при малой энергии движение близко к гармоническим колебаниям в потенциальной яме. Вплоть до сепаратриссы дрейф и диффузия дислокаций идёт посредством туннелирования.  Когда колебательной энергии дислокации хватает для преодоления барьера Пайерлса-Набарро, то начинается новый режим со сдвиговой деформацией в плоскости скольжения дислокаций.  Если приложено внешнее напряжение сдвига, то меняется вероятность движения вдоль и против силы. Пока энергия дислокаций невелика, мала и скорость перемещения дислокаций по допустимым состояниям.  В этом случае расчёт энтропии сводится к комбинаторике, с учётом вероятности расположения дислокаций по решётке. Строгий анализ предполагает учёт колебательных степеней свободы, возбуждаемых с ростом внутренней энергии в дислокациях.

Наш анализ вызван попыткой разобраться со странностями в определении энтропии, приведённой для подвижной дислокации в \cite{GN1}. На эволюцию числа дислокаций влияют процессы их рождения и исчезновения, но это не важно при расчёте энтропии. Теплопередачу от дислокаций к решётке считаем медленной.

     \section{Энтропия как результат подсчёта числа допустимых состояний системы}

\begin{figure}[t]
\begin{center}
  \includegraphics[scale = 0.9]{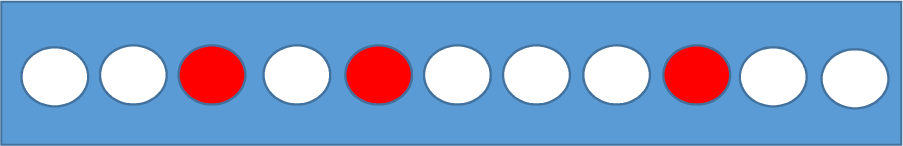}
\caption{ Возможное положение дислокаций и их реальное размещение. Места, закрашенные цветом, показывают реализованные размещения, прочие - возможные.  }% Give a unique label
\label{fig:r1}
\end{center}
\end{figure}

Чем больше у системы допустимых состояний, тем больше энтропия \cite{Kit}.  Количество дислокаций обычно конечно и невелико по сравнению количеством возможных позиций их размещения. Задача о вычислении числа допустимых состояний системы эквивалентна проблеме размещения заданного числа $k$ шаров в коробку с большим числом мест $n$ (см. рисунок \ref{fig:r1}), и определяется числом сочетаний без учёта порядка $M = C_n^k = \frac{{n!}}{{(n - k)!k!}}$. Здесь $n$ -полное количество мест, $k$ -число дислокаций. Поскольку как занятые, так и свободные позиции ничем не отличаются, то в знаменателе стоят соответствующие делители числа перестановок.  Переходя к энтропии, необходимо взять логарифм числа состояний, который состоит из трёх факториалов. Логарифм факториала берётся по асимптотической формуле $\ln n! = \frac{1}{2}\ln (2\pi n) + n\ln \frac{n}{e}$.

%%  \includegraphics[bb=1cm 1cm 15cm 3cm]{pic7con.eps} %%{doc4.pdf}
%%\vspace{1.cm}       % Give the correct figure height in cm

С учётом этого
$$\begin{array}{l}
\ln \frac{{n!}}{{(n - k)!}} = \frac{1}{2}\ln (\frac{n}{{n - k}}) + n\ln \frac{n}{e} - (n - k)\ln \frac{{n - k}}{e} = k\ln \frac{{n - k}}{e} + (\frac{1}{2} + n)\ln (\frac{1}{{1 - k/n}}) = \\
 = k\ln \frac{{n - k}}{e} + (\frac{1}{2} + n)\frac{k}{n} = k\ln (n - k)\\
\ln \frac{{n!}}{{(n - k)!k!}} = k\ln \frac{{e(n - k)}}{k} - \frac{1}{2}\ln (2\pi k)
\end{array}$$
Тогда на одну дислокацию приходится энтропия, равная приблизительно
\begin{equation}\label{eq:nn2}
{s_0} = 1 + \ln {n_1},
\end{equation}
где ${n_1} = n/k - 1$ число возможных состояний в пересчёте на одну дислокацию.
Данный результат сопоставим с двумерным случаем \cite{t6}, ${S_0} = {A_0}\rho \ln \frac{e}{{{b^2}\rho }}$, $b$ - шаг решётки, под знаком логарифма стоит отношение числа дислокаций к числу позиций ${b^2}\rho  = k/n$, а множитель ${A_0}\rho $ - это число дислокаций на площади среза $n{b^2}(k/n{b^2}) = k$. При $n >> k$  результаты совпадают.
Рассмотрим более сложную ситуацию, когда вероятность каждой позиции меняется из-за поля взаимодействия с соседними дислокациями. Поскольку энергия, запасённая в дислокации зависит от ее радиуса, то изменение плотности дислокаций приводит к изменению вероятности возможных состояний. Для задания вероятности данной позиции дислокации будем использовать распределение Гиббса. Поскольку речь идёт о локальном тепловом равновесии, то роль температуры будет осуществлять некоторая локальная температура дислокаций. Далее упростим задачу, считая модель одномерной. Считаем на некотором участке цепочки плотность дислокаций постоянной. Тогда локально энтропия, внутренняя и свободная энергия пропорциональны числу дислокаций $k$. Теперь перейдём к расчёту энтропии одной дислокации на выбранном участке, заморозив позиции соседних дислокаций в среднем положении. Среднее расстояние до соседей в шагах решётки равно ${n_1}$. Число возможных позиций выделенной дислокации относительно правого соседа равно ${n_1}$. Шаг к правому соседу приводит к удалению от левого соседа. Для задания энергии единицы длины дислокации\cite{MCh} обычно пользуются формулой:
\begin{equation}\label{eq:nn3}
{H_ \bot }(k) = \frac{{G{b^2}}}{{10}} + \frac{{G{b^2}}}{{4\pi (1 - \nu)}}\ln \frac{R}{{5b}}
\end{equation}
где  $R = {\rho ^{ - 1/2}}$ - средний радиус площадки вокруг дислокации, где присутствует поле напряжений только от одной дислокации (здесь $\rho $ - плотность дислокаций, $G$ - модуль сдвига), для краевых дислокаций ($\nu = \frac{1}{3} - {\rm{Poisson\; ratio}}$).
При плотности $\rho  = 4 \cdot {10^{10}}/{\rm{c{m^2}}}$ радиус равен $5 \cdot {10^{ - 2}}{\rm\mu m} = 200b$ и погонная энергия равна ${H_ \bot } \approx G{b^2}/2 = 1.3 \cdot {10^{ - 9}}{\rm{J/m = 2eV/b}}$ (во втором случае -- в пересчёте на атомную плоскость).
Введём обозначение $\Delta  = \frac{{G{b^3}}}{{4\pi (1 - v)}} = 0.47eV$ и перейдём к анализу одномерной ситуации. Энергия, накопленная в поле напряжений при переносе правой ближней дислокации с ${R_\infty }$ до ${R_{+}}$ равна ${U_{ +}} = (\Delta /2)\ln ({R_\infty }/{R_{+} })$. Если до левого соседа расстояние ${R_{-} }$, а среднее расстояние, полученное через плотность дислокаций, есть $\bar R$, то прибавка энергии из-за смещения с наиболее вероятной позиции определяется как
\begin{equation}
{U_p} = {U_+} + {U_-} - 2{U_{\bar R}} = \frac{\Delta }{2}\ln \frac{{{{\bar R}^2}}}{{{R_+}{R_-}}} = \frac{\Delta }{2}\ln \frac{{n_1^2}}{{({n_1} + p)({n_1} - p)}}
\end{equation}
Здесь $\overline R  = {n_1}b = (n/k - 1)b$ представлено через среднее число мест, в которых возможно найти дислокацию. Тогда как фактические координаты пробегают позиции ${R_\pm } = b({n_1} \pm p)$. Вероятность состояния со смещением дислокации на $p$ позиций от симметричного состояния определяется формулой Гиббса $p = \exp ( - {U_p}/T)$, тогда свободная энергия равна
\begin{equation}\label{eq:nn5}
f =  - T\ln (\sum\limits_p^{} {{e^{ - {U_p}/T}}} ) =  - T\ln (\sum\limits_{p = 0}^{{n_1}} {\exp (\frac{\Delta }{{2T}}\{ \ln (1 + \frac{p}{{{n_1}}}) + \ln (1 - \frac{p}{{{n_1}}})\} } )
\end{equation}
В (\ref{eq:nn5}) свободная энергия вычислена через статистическую сумму состояний выделенной дислокации, когда позиции остальных заморожены. От суммирования можно перейти к интегрированию в пределе, когда ${n_1} \gg 1$. Для удобства обозначим показатель степени через $\alpha  = \frac{\Delta }{{2T}}$, температура измеряется в электрон-вольтах (при комнатной температуре $T = 0.025eV$). Тогда результат выражается через бета-функцию:
\[\begin{array}{l}
f =  - T\ln (\frac{1}{2}\sum\limits_{p =  - {n_1}}^{{n_1}} {{e^{\alpha (\ln (1 + \frac{p}{{{n_1}}}) + \ln (1 - \frac{p}{{{n_1}}}))}}} ) =  - T\ln \{ {2^{2\alpha }}{n_1}\int\limits_0^1 {{x^\alpha }} {(1 - x)^\alpha }dx\} \\
 =  - T\ln \{ {2^{2\alpha }}{n_1}\frac{{{\Gamma ^2}(1 + \alpha )}}{{\Gamma (2 + 2\alpha )}}\}
\end{array}\]
Напомним, в одномерной цепочке два соседа, поэтому в показателе экспоненты два слагаемых, из-за чётности по $p$ область суммирования делаем симметричной, поделив результат пополам. Величина свободной энергии выражается через бета-функцию, здесь от суммирования можно перейти к интегрированию (так как ${n_1} >  > 1$ ). Энтропия равна ${s_1} =  - \frac{{\partial f}}{{\partial T}}$, а поскольку $\frac{{\partial (2\alpha T)}}{{\partial T}} = 0$, множитель ${2^{2\alpha }}$  не даёт вклад в конечный результат.  Выражение для энтропии одной дислокации имеет следующий вид:

\begin{equation}\label{eq:nn6}
{s_1} = \ln {n_1} - \frac{\partial }{{\partial T}}[T\ln \frac{{\Gamma (2 + 2\alpha )}}{{{\Gamma ^2}(1 + \alpha )}}] \approx \ln {n_1} - \frac{\partial }{{\partial T}}[T\ln \frac{{(1 + 2\alpha )!}}{{{{(\alpha !)}^2}}}]
\end{equation}

Здесь первое слагаемое - конфигурационная энтропия, которая совпадает с ${s_0} - 1$, из (\ref{eq:nn2}). А второе слагаемое из (\ref{eq:nn6}) возникло из-за изменения поля вероятности под действием поля напряжений вокруг дислокаций. При низких температурах $\alpha  = \frac{\Delta }{{2T}} >  > 1$ результат сводится к элементарным функциям:
\begin{equation}\label{eq:nn7}
{s_{app}} = \ln {n_1} - \frac{\partial }{{\partial T}}[T\ln (\frac{{1 + 2\alpha }}{{\sqrt {\pi \alpha } }})] = {s_0} - \frac{1}{2} + \frac{1}{2}\ln (\frac{{\pi T}}{{2\Delta }})
\end{equation}
Логарифм в (\ref{eq:nn7}) отрицателен, что свидетельствует об падении вероятности выбора многих доступных состояний в поле напряжений, окружающих дислокацию, и поэтому средняя энтропия меньше, чем ${s_0}$. Если $n/k = 2000$, то ${s_0}=1+\ln\frac{{2L}}{{kb}} = 1 + \ln \frac{n}{k} = 8.6$. А при нормальной температуре $\frac{\Delta }{T} = 20$, энтропия есть $s = \frac{1}{2} + \ln{n_1} + \frac{1}{2}\ln \frac{{\pi T}}{{2\Delta }} = {s_0} - 1.8 = 6.8$.

\begin{figure}[ht]
\begin{center}
\includegraphics[scale=0.9]{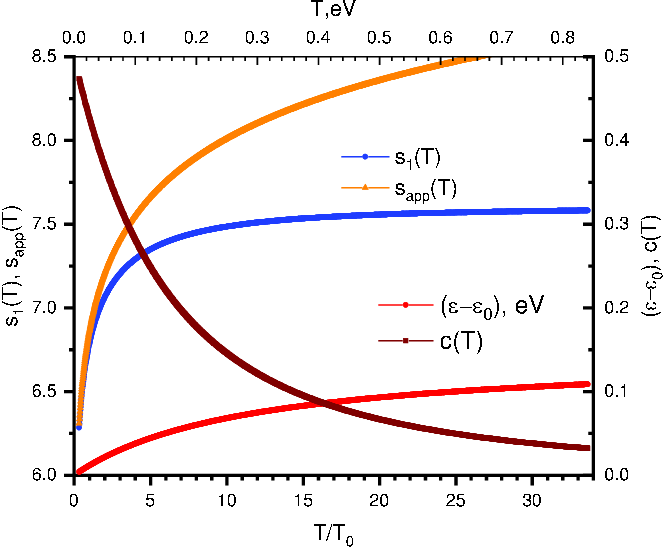}
%%  \includegraphics[bb=1cm 1cm 14cm 9cm]{pic2con.eps} %%{doc4.pdf}
%%\vspace{1.cm}       % Give the correct figure height in cm
\caption{ Энтропия (ось слева) и энергия (ось справа) на одну дислокацию при разных температурах. Кривая $s_1(T)$ (синяя) соответствует энтропии по формуле(\ref{eq:nn8}), вторая ${s_{app}}(T)$ - аппроксимации (\ref{eq:nn7})-(оранжевая) . Температура и энергия(\ref{eq:nn9})-(красная линяя) - измеряются в электронвольтах, теплоёмкость(коричневая) - безразмерная. Число вакансий на дислокацию взято $n/k = 2000$.}
\label{fig:02}       % Give a unique label
\end{center}
\end{figure}

Сравнивая с результатом \cite{t7}, когда на 200 атомов приходиться 2 дислокации в двух измерениях, в одном измерении отношение числа дислокаций к числу узлов решётки 1:10, тогда сдвиг энтропии $\Delta s$ в пересчёте на один атом решётки даёт величину $1.8 \times 0.1 = 0.18$. Это число одного порядка с результатом \cite{t7}. Однако, в нашем анализе учитываются обе части энтропии: конфигурационная энтропии ${s_0}$ и термодинамическая энтропия.
Также приведём точный результат ($\alpha  = \frac{\Delta }{{2T}}$) через пси и гамма-функции (см. рисунок \ref{fig:02}):
\begin{equation}\label{eq:nn8}
s_1 = \ln {n_1(\rho )} - \ln \frac{{\Gamma (2 + 2\alpha )}}{{{\Gamma ^2}(1 + \alpha )}} + 2\alpha \{ \psi (2 + 2\alpha ) - \psi (1 + \alpha )\}
\end{equation}

Изменение внутренней энергии $\varepsilon  = f + Ts$, отсчитанного от уровня запасённой энергии ${\varepsilon _0}$ с наиболее вероятным расстоянием между дислокациями, задаётся формулой:
\begin{equation} \label{eq:nn9}
\varepsilon  - {\varepsilon _0} = \Delta \{ \psi (2 + 2\alpha ) - \psi (1 + \alpha ) - ln2\}  = \frac{\Delta }{2}\{ \psi (1 + \alpha  + \frac{1}{2}) - \psi (1 + \alpha )\}
\end{equation}

Если плотность дислокаций переменная, как это показано выше в случае с препятствиями на обоих концах кристалла\cite{LL7}, то средняя энтропия, приведённая на одну дислокацию, есть
\begin{equation}\label{eq:n27}
\begin{array}{l}
s = 1 -  < \ln \rho  >  = 1 - \frac{1}{{kb}} \int\limits_{ - L}^L {\frac{1}{{\pi \sqrt {{L^2} - {\xi ^2}} }}} \ln \frac{{kb}}{{\pi \sqrt {{L^2} - {\xi ^2}} }}d\xi  = \\
 = 1 - \frac{2}{\pi }\int\limits_0^{\pi /2} {\ln \frac{{kb}}{{\pi L\cos t}}} dt = 1 + \ln \frac{{\pi L}}{{kb}} + \frac{2}{\pi }\int\limits_0^{\pi /2} {\ln \cos t} dt = {s_0} + \ln \frac{\pi }{2} - \ln 2 = {s_0} - 0.24
\end{array}   	
\end{equation}

Поскольку плотность дислокаций меняется при приложенном давлении как \cite{LL7}
\begin{equation}\label{eq:n28}
\rho  = \frac{1}{{\pi \sqrt {{L^2} - {x^2}} }}(\frac{{p(x)}}{D}x + kb),
\end{equation}
мы изучим случай, когда приложенное напряжение $p(x) \equiv {p}$ компенсируется перераспределением дислокаций (\ref{eq:n28}). Максимальное напряжение в модели распределенных дислокаций равно ${p_{\max }} = kbD/L$ ($k$ -число дислокаций на длине зерна, $b$ -длина вектора Бюргерса). Энтропия, как функция $p(x)$ может быть сведена к формуле:
\begin{equation}\label{eq:n29}
s(p)\left| {_{p = {p_{\max }}}} \right. = \ln \frac{{\pi L}}{{kb}} + \sqrt {1 - {{(\frac{p}{{{p_{\max }}}})}^2}}  - \ln (1 + \sqrt {1 - {{(\frac{{{p}}}{{{p_{\max }}}})}^2}} ) = {s_0} + \ln \frac{\pi }{2} - 1 = {s_0} - 0.55	
\end{equation}

Перераспределение дислокаций из-за наложенного внешнего давления понижает среднюю по системе энтропию. Отметим отдельно случай высоких температур. Разлагая экспоненту в статистической сумме в ряд с точностью до вторых порядков, получим
\begin{equation}\label{eq:n30}
S =  - \frac{{\partial F}}{{\partial T}} = \ln M - \frac{{ < {H_k}^2 >  -  < {H_k}{ > ^2}}}{{2{T^2}}}
\end{equation} 	
Здесь $\ln M = k\ln e{n_1}$, ${n_1}$ -число узлов решетки в пересчете на одну дислокацию: ${n_1} = n/k - 1$, усреднение энергии выполняется по всем возможным состояниям системы дислокаций. В правой части (\ref{eq:n30}) с точностью до знака и константы  $\frac{1}{{2{T^2}}}$  стоит дисперсия полной энергии. Таким образом, $\ln M$ - верхняя грань величины энтропии. Здесь кроется коренное отличие от энтропии идеального газа - поскольку с ростом температуры энтропия насыщается. Поэтому график $s_1(T)$ (см.  Рисунок 2)  заметно отличается от аппроксимации по обычной зависимости от температуры (\ref{eq:nn7}).

Чтобы перейти к полному описанию энтропии дислокаций, необходимо ввести в описание периодический потенциал вида Пайерлса-Набарро. Полный потенциал - это сумма периодического потенциала и потенциала, образованного полем напряжений дислокаций. Из-за структуры статистической суммы энтропия входит аддитивно от каждого вклада в свободную энергию. Периодический потенциал даст новое слагаемое в полную энтропию. По-прежнему считаем, что дислокации распределены с некоторой эффективной температурой, которая задаёт вероятность распределения по формуле Гиббса. Однако, в отличие от расчёта по теории гармонического маятника для атомов самой решётки, колебания дислокаций в области пластичности уходят в режим осцилляций далёкий от гармонического.  Поэтому уровни энергии лежат не эквидистантно, более того они сгущаются к точке сепаратриссы между локализованным и инфинитным режимами движения. В анализе термодинамики твёрдого тела, основанном на гармоническом осцилляторе в области высоких температур, мерой числа состояний служит отношение температуры к средней основной частоте осциллятора - $s = \ln \frac{{eT}}{{\hbar \bar \omega }}$ - то есть число возбуждённых колебательных степеней свободы. Когда уровни энергии сгущаются при приближении к сепаратриссе в режим с инфинитным движением, то их вовлечение в динамику приведёт резкому росту степеней свободы колебательных мод. Поэтому, рост энтропии будет усилен вблизи точки фазового перехода.

Отметим некоторые особенности, которые возникают при изучении дислокаций в системах большей размерности. В работе \cite{SBS} обсуждается множество локальных минимумов энергии взаимодействия ансамбля краевых дислокаций в двумерном случае. В противоположность данным выводам в курсе \cite{LL7} методами ТФКП показано (в рамках модели непрерывной плотности дислокаций), что решение о минимуме энергии имеет единственное стационарное решение в одномерном случае. Причём характер решения задаётся условием на границах зерна. Численное решение в одномерной модели дискретных дислокаций также подтверждает выводы аналитической теории. Отметим, что в статье\cite{SBS} выбрано дважды периодическое граничное условие для функции свободной энергии в области определения дислокации, которое может противоречить с физическими условиями на границе зерна. В работах \cite{SBS,Be} плотность энтропии определена как логарифм числа p-состояний на единицу объёма для заданной энергии системы; ${\bf p}$ - это средняя поляризация. В \cite{Be} приращение тензора поляризации дислокаций эквивалентно инкременту тензора пластической деформации. Грубо говоря, поляризация дислокаций является частью пластической деформации, которая, однако, не зависит от истории. Данная схема отличается от принятой в следующем разделе. У нас равновесное поле напряжений - это константа по зерну и есть функция краевых условий. Задание внутренней энергии системы полностью определяет число доступных состояний, которое эквивалентно числу колебательных мод, возбуждаемых на этом уровне энергии. Для описания уравнения состояния дислокации мы используем модельный потенциал, эквивалентный потенциалу Пайерлса-Набарро. Динамика пластических деформаций реализуется, когда внутренняя энергия превышает порог. В этом случае возникает поступательное движение, сопровождаемое колебательным, а скорость зависит от полной энергии. Поступательное движение - это по сути дополнительная к колебательной степень свободы. В нашем случае независимость от истории выполняется автоматически, так как число состояний - функция только полной внутренней энергии, не зависящей от предыстории деформаций.

\section{Энтропия дислокации в рамках модели физического маятника}

В этом разделе статистическую сумму состояния будем задавать на основе распределения Гиббса, то есть вероятность состояния определяется его полной внутренней энергией и средней температурой по группе дислокаций. Для использования такого определения вероятности состояния есть определённые основания, связанные с возникновением локальной равновесности по температуре. Однако, система - неравновесная по отношению к числу дислокаций из-за большой энергии их возникновения. Дислокации могут туннелировать на новые (более вероятные) позиции, но скорости этих процессов малы при обычных температурах. И кроме того есть вопросы: выполняется ли эргодичность (когда среднее по ансамблю равно среднему по времени); является ли динамическая система стохастичной (применимость статистической теории). В \cite{Zasl} указывается, что свойство эргодичности, введённое ещё в работах Гиббса, опирается на теорему Пуанкаре о возвратах. Но требуется условие перемешивания для возникновения условий конечности времени релаксации системы к статистически равновесной. Сама теорема Пуанкаре не имеет никакого отношения к появлению стохастичности. Анализ, проведённый Крыловым \cite{Kryl}, показал, что в основе появления статистических законов лежит не свойство эргодичности, а свойство перемешивания. В самих уравнениях динамики системы не содержится механизм огрубления, даже если динамика квантовая. Заметим, что из условия перемешивания автоматически следует и свойство эргодичности. Современные достижения эргодической теории показали, что для появления статистических свойств вовсе не обязательно, чтобы система состояла из большого числа степеней свободы. Эти свойства могут быть, начиная с системы двух частиц. Статистические свойства, включая энтропию, задаются на фазовом пространстве. Больцман ввёл определение энтропии через функцию распределения. Однако, если функция распределения удовлетворяет уравнению Лиувилля, то энтропия, определённая по функции распределения, не может зависеть от времени. Замена точной функции распределения на огрублённую - крупнозернистую - приводит к необратимости, и энтропия начинает зависеть от времени, а система движется к равновесной. Вблизи равновесия состояния системы приблизительно равномерно распределяются в некоторой области фазового пространства размером со статистический вес этого состояния, а динамическая система ведёт себя как случайная.
Хотя наш случай и тяготеет к неравновесному, однако к такому, которое \cite{t4} можно рассмотреть посредством технологии неполного равновесия. Выбором надлежащего разбиения системы на части, которые на некотором интервале времени  будут частично равновесными, можно найти энтропию системы как сумму этих изменяющихся во-времени частей. В рамках данного представления функция распределения будет иметь вид распределения Гиббса также, как и в равновесном случае, а возможная зависимость от координат и времени становится неявной - через набор независимых термодинамических переменных, таких как $T(x)$. Теперь перейдём к учёту вклада колебательных мод в потенциале Пайерлса-Набарро.

В этом разделе немного изменим классический подход из статистической физики для твёрдых тел при температуре выше температуры Дебая. Для задания  термодинамических величин часто вместо температуры Дебая используется связанная с ней средняя частота ансамбля атомов внутри зерна. Согласно определению,  средне-геометрической частотой системы атомов с $3N$ колебательными степенями свободы будет величина, полученная суммированием логарифмов $\ln \bar \omega  = \frac{1}{{3N}}\sum\limits_\alpha  {\ln {\omega _\alpha }}$. Энтропия на один атом равна $s = \ln \frac{{eT}}{{\hbar \bar \omega }}$; при выводе   используется статистическая сумма из вероятностей заполнения по уровням математического маятника. При замене объекта физики с колебательных мод атомов решётки на колебания одномерных дефектов - дислокаций, модель математического маятника трансформируется в физический маятник.  При этом потенциал меняет форму с параболического на периодический потенциал Пайерлса-Набарро. Одновременно с изменением формы потенциала меняется и расстояние между уровнями по мере подъёма со дна ямы. Из-за конечности энергетического барьера между ямами, расположение которых совпадает с положением узлов решётки, внешнее поле напряжений выше некоторого порога приводит к скольжению дислокаций. Итак, для описания динамики дислокаций подходит модель физического маятника с гамильтонианом	
\begin{equation}\label{eq:n1}
H = \frac{1}{2}{\dot x^2} - \omega _0^2\cos x
\end{equation}
Здесь ${\omega _0} = \sqrt {\frac{K}{m}}  = \sqrt {\frac{{\mu b}}{m}}  = 1.09 \cdot {10^{13}}{{\rm{s}}^{ - 1}}$, модуль сдвига $\mu  = 37\Gamma {\rm{Pa}}$, вектор Бюргерса $b = 2.57 \cdot {10^{ - 10}}{\rm{m}}$, масса единицы длины $b$ дислокации взята $m = 8 \cdot {10^{ - 26}}{\rm{kg}}$. Отметим, что динамическая масса дислокации, согласно теории \cite{s17}, не локальна, однако пропорциональна плотности среды. Нелокальность приводит к логарифмическому множителю от отношения радиуса кривизны дислокации к радиусу ее сердцевины (случай замкнутой петли дислокации). Влияние этого множителя может увеличить массу 4-5 раз. Размерный гамильтониан равен ${\lambda _\dag }H$, где $\lambda  = {\lambda _\dag }\omega _0^2 = \frac{{m{b^2}}}{{{\pi ^2}}}\omega _0^2 = \frac{{\mu {b^3}}}{{{\pi ^2}}} = 0.4{\rm{eV}}$ и $\mu  = \frac{{2\lambda }}{{\hbar {\omega _0}}} = {\rm{114}}$. Переход к безразмерной координате осуществляется нормировкой на вектор Бюргерса, времени - умножением на частоту ${\omega _0}$. Безразмерный гамильтониан в новых переменных имеет вид
\begin{equation}\label{eq:n2}
\hat H = \frac{1}{2}{\dot x^2} - \cos x
\end{equation}
Если мерить энергию от дна ямы, то следует ввести новый параметр $2{\kappa ^2} = (\hat H + 1)$, с учетом вышесказанного запасенная в колебаниях энергия равна $2\lambda {\kappa ^2}$. Когда колебания локализованные, то параметр ${\kappa ^2} < 1$. В противном случае, параметр ${\kappa ^2} > 1$. На границе этих областей располагается сепаратрисса, отмечающая фазовый переход между двумя различными типами колебаний дислокации. В первом случае решение уравнения (\ref{eq:n2}) можно искать в виде эллиптических функций от безразмерных величин $x = {x_\dag }/b,t = {\omega _0}{t_\dag }$:	
\begin{equation}\label{eq:n3}
\begin{array}{l}
x = 2\arcsin (\kappa \sin \xi )\\
\dot x = 2\kappa {\mathop{\rm cn}\nolimits} (t,\kappa )\\
\xi  = {\mathop{\rm am}\nolimits} (t,\kappa ) ={\rm arcsin}({\rm sn}(t,\kappa ))\\
\frac{{2I}}{\mu } = \frac{1}{\pi }\int\limits_{ - x(K)}^{x(K)} {\dot xdx = } \frac{{4{\kappa ^2}}}{\pi }\int\limits_0^{2K} {{{{\mathop{\rm cn}\nolimits} }^2}(t,\kappa )dt = } \frac{8}{\pi }(E - (1 - {\kappa ^2})K)
\end{array}
\end{equation}
Частота колебаний зависит от энергии, запасенной в системе физического осциллятора
\begin{equation}\label{eq:n4}
\begin{array}{l}
\omega ({\hat H}) = \frac{\pi }{{2K(\kappa )}},\\
K(\kappa ) = \int\limits_0^{\pi /2} {\frac{{d\xi }}{{\sqrt {1 - {\kappa ^2}{{\sin }^2}\xi } }}},
\end{array}\end{equation}
где $K$ - полный эллиптический интеграл первого вида. Данная частота определяет разбиение фазового пространства в соответствие с правилом квантования Бора-Зоммерфельда \cite{LL3}. Условие квантования Бора-Зоммерфельда задаётся формулой, где интеграл берётся по полному периоду классического движения $I = \frac{1}{{2\pi \hbar }}\oint {pdx}  = n + \frac{1}{2}$, где $I$ - адиабатический инвариант движения (действие) и $I = \int\limits_0^{{\kappa ^2}} {\frac{{dI}}{{d{\kappa ^2}}}} d{\kappa ^2} = \mu \int\limits_0^{{\kappa ^2}} {\frac{{2K(\kappa )}}{\pi }d{\kappa ^2}}  = n + \frac{1}{2}$.  Уровни с учётом безразмерного фактора $\mu  = \frac{{2\lambda }}{{\hbar {\omega _0}}}$, возникшего при приведении к безразмерным величинам, располагаются по правилу квантования. Отсчитывая от дна ямы, энергию уровня связываем с частотой кванта рекуррентным соотношением:
\begin{equation}\label{eq:n5}
\frac{{\Delta {\varepsilon _n}}}{{\hbar {\omega _0}}} = \frac{\mu }{2}(\kappa _n^2 - \kappa _{n - 1}^2) = {[\frac{1}{2}({\omega ^{ - 1}}(\kappa _n^2) + {\omega ^{ - 1}}(\kappa _{n - 1}^2))]^{ - 1}} \approx \omega (\kappa _n^2)
\end{equation}
В пределе $\kappa  \to 1$ уровни энергии сгущаются. Используя асимптотику $I$ при ${\kappa _n} \to 1$, можно получить оценку для числа уровней ввиду $\frac{{dI}}{{d{\kappa _n}^2}} \to \frac{\mu }{\pi }\ln \frac{1}{{(1 - {\kappa _n}^2)}}$. Заменим $\kappa _{\max  + 1}^2 \to 1$, тогда
\begin{equation}\label{eq:n6}
(1 - {\kappa _{\max }}^2)\ln \frac{1}{{(1 - {\kappa _{\max }}^2)}} = \pi /\mu \end{equation}

Для $\pi /2\mu  = 1.37 \cdot {10^{ - 2}}$ величина $(1 - {\kappa _{\max }}^2) = 5. \cdot {10^{ - 3}}$. Для этого уровня частота колебаний падает вдвое $\omega  = \frac{\pi }{{2K}} = \frac{\pi }{{\ln {{(1 - {\kappa _{\max }}^2)}^{ - 1}}}} = 0.59$. Число гармоник порядка $n = \frac{{\mu \kappa _n^2}}{{\bar \omega (\kappa _n^2)}} \to \frac{\mu }{{\bar \omega }} = 143$. Расчёт даёт 145 уровней. Ёмкость фазового пространства увеличивается при приближении к сепаратриссе.

Если система имеет высокую температуру, то ряд дислокаций может пересечь границу фазового перехода и перестать быть локализованными. Этому соответствует условие ${\kappa ^2} > 1$ .  В этом случает энергия дислокации может быть представлена из двух частей - кинетическая энергия равномерного движения и колебательная плюс потенциальная части энергии, отвечающая осцилляциям в яме. Решение (\ref{eq:n2}) выше сепаратриссы можно описать также в эллиптических функциях:
\begin{equation}\label{eq:n10}
\begin{array}{l}
x = 2\arcsin ({\mathop{\rm sn}\nolimits} (\kappa t,{\kappa ^{ - 1}})) = 2{\mathop{\rm am}\nolimits} (\kappa t,\kappa )\\
\dot x = 2\kappa {\mathop{\rm dn}\nolimits} (\kappa t,{\kappa ^{ - 1}})
\\\frac{{2I}}{\mu } = \frac{1}{\pi }\int\limits_0^{x(2K)} {\dot xdx = } \frac{4}{\pi }\int\limits_0^{2K} {\kappa {{{\mathop{\rm dn}\nolimits} }^2}(z,{\kappa ^{ - 1}})dz = } \frac{{8\kappa }}{\pi }E({\kappa ^{ - 1}})
\end{array}\end{equation}

Шкала времени изменилась на множитель $\kappa $. Интегральный инвариант Пуанкаре \cite{Aiz} задан в расширенном $2n + 1$ - мерном фазовом пространстве $(x,\dot x,t)$ на одно-параметрическом контуре интегрирования, охватывающем один период колебаний (разность значений на концах контура $\Delta x = 2\pi ,\Delta \dot x = 0$, по $x$ координате кольцо по модулю $2\pi $), и не зависит от выбора начального момента времени. В квазиклассическом приближении волновые функций связаны с действием; из их ортогональности между собой следует, что действие равно натуральному числу $I = n$. Данное условие означает, что выше сепаратриссы потенциал - безотражательный. Частота колебаний зависит от энергии по закону
\begin{equation}\label{eq:n11}
\begin{array}{l}
{\omega _{{k^{ - 1}}}} = \frac{{\kappa \pi }}{{2K({\kappa ^{ - 1}})}}\\
K({\kappa ^{ - 1}}) = \int\limits_0^{\pi /2} {\frac{{d\xi }}{{\sqrt {1 - {\kappa ^{ - 2}}{{\sin }^2}\xi } }}}  = \frac{\pi }{2}\,{F_2}_1(\frac{1}{2},\frac{1}{2};1;{\kappa ^{ - 2}})\\
E({\kappa ^{ - 1}}) = \int\limits_0^{\pi /2} {\sqrt {1 - {\kappa ^{ - 2}}{{\sin }^2}\xi } d\xi  = } \frac{\pi }{2}\,{F_2}_1( - \frac{1}{2},\frac{1}{2};1;{\kappa ^{ - 2}})
\end{array}\end{equation}

Функция $E({\kappa ^{ - 1}})$ монотонно растёт от 1 до ${\pi/2}$, когда аргумент ${\kappa ^{ - 1}}$ изменяется от 1 до 0. Также растёт $\kappa$ на том же промежутке. Минимальное значение действия достигается при ${\kappa ^{ - 1}} = 1 - \varepsilon ,\varepsilon  > 0$, при этом действие примерно равно $I({\kappa ^{-1}})=4\mu/\pi=\mathord{\buildrel{\lower3pt\hbox{$\scriptscriptstyle\frown$}}
\over n}  = 145$. Вблизи сепаратриссы уровни располагаются плотнее, на асимптотике, когда ${\kappa ^{ - 1}} \to 0$  уровни расположены по закону $2\mu {\kappa _n} = n$. Средняя скорость дрейфа находится из разложения функции $\dot x$ в ряд по гармоникам, откуда $2\kappa \overline {{\mathop{\rm dn}\nolimits} (\kappa t,{\kappa ^{ - 1}})}  = 2{\omega _{{k^{ - 1}}}}$. Для ${\kappa _n} > 2{\kappa _0}$ энергия поступательного движения превалирует над колебательной, а частота ${\omega _{{k^{ - 1}}}} \to {\kappa _n}$. Каждый период колебаний сопровождается перескоком дислокации в соседнюю яму, то есть средняя скорость квантована с помощью единственного параметра ${\omega _{{k^{ - 1}}}}$.

Рассмотрим статистическую сумму данного набора уровней в системе из $N$ независимых дислокаций. Независимость дислокаций или стохастичность их состояний приводит к факторизации полной статистической суммы по единичным статистическим суммам $Z$.
\begin{equation}\label{eq:n21}
\hat Z = {Z^N}
\end{equation}
\begin{equation}\label{eq:n22}
\frac{F}{N} =  - T\ln \left( {\sum\limits_n^{} {{e^{ - 2\lambda \kappa _n^2/T}}} } \right) =  - T\ln Z(T)
\end{equation}
Сумма идёт по состояниям как с ${\kappa _n} < 1$, так и по состояниям выше сепаратриссы с ${\kappa _n} > 1$.  А энтропия в пересчёте на одну дислокацию равна
\begin{equation}\label{eq:n23}
\begin{array}{l}
s = \frac{S}{N} =  - \frac{1}{N}\frac{{\partial F}}{{\partial T}} = \ln Z(T) + T\frac{\partial }{{\partial T}}\ln Z(T)\\
{s_{sp}} =  - \ln (1 - {e^{ - \hbar \omega /T}}) + \frac{{\hbar \omega /T}}{{{e^{\hbar \omega /T}} - 1}}
\end{array}
\end{equation}
Вторая формула энтропии ${s_{sp}}$ в (\ref{eq:n23}) приведена в приближении колебаний квантового математического маятника \cite{t4}, когда его частота равна частоте колебаний на нижнем уровне физического маятника. Внутренняя энергия равна $E = F + ST$, теплоёмкость - производная по температуре; значения обоих величин вычисляются по формулам(\ref{eq:n24}), ${c_{sp}}$ - формула для  сравнения с простым квантовым осциллятором.
\begin{equation}\label{eq:n24}
\begin{array}{l}
u = \frac{E}{N} = {T^2}\frac{{\partial \ln (Z)}}{{\partial T}},\;c = \frac{{\partial u}}{{\partial T}}\\
{c_{sp}} = {(\frac{{\hbar \omega }}{T})^2}\frac{{{e^{\hbar \omega /T}}}}{{{{({e^{\hbar \omega /T}} - 1)}^2}}} \to 1
\end{array}
\end{equation}
Энергия запасена во внутренних степенях свободы и учитывает сгущение уровней вблизи фазового перехода. В данном анализе рассмотрены только дискретные уровни для локализованных состояний и дискретные уровни в режиме с инфинитным движением.

\section{Численные результаты}

Дислокации образуют собой систему взаимодействующих дефектов через поле напряжений. Для определённости будем считать, что размеры монокристалла порядка 50 микрон вдоль координаты $x$.  На этом отрезке расположены $k=100$ одинаково ориентированных дислокаций, которые находятся в равновесии. Напряжение поля дислокаций задаётся формулой
\begin{equation}\label{eq:n25}
\begin{array}{l}
{\sigma _{xy}}({x_i}) =  - Db\sum\limits_{j = 1}^{k} {\frac{1}{{{x_j} - {x_i}}}} ,i = 2,...,k-1\\
{x_1} = 0,{x_{100}} = 50
\end{array}
\end{equation}

Константа $D$ связана с модулем сдвига $D = \mu /(2\pi (1 - \nu))$.  Решение(\ref{eq:n25}) находится численно, как стационар системы уравнений вида ${\ddot x_i} + \gamma {\dot x_i} = \sigma ({x_i}) + p$, и заданием нормы, как суммы абсолютных значений компонент вектора $\sigma ({x_i}) + p$. $p$ - давление внутри зерна, которое компенсируется напряжением поля дислокаций. Примерное расположение дислокаций описывается формулой, полученной аналитически в модели непрерывного распределения дислокаций при нулевом давлении. $\rho (x)$ - локальная плотность дислокаций \cite{LL7}. Здесь $\int\limits_{ - L}^{{x_i}} {\rho (x)dx = ib}, i\le k $

\begin{equation}\label{eq:n26}
\rho (x) = \frac{{bk}}{{\pi \sqrt {{L^2} - {x^2}} }},L = 25{\rm{\mu m}} ,k = 100
\end{equation}

Положение равновесных дислокаций в первом приближении управляется формулой
\begin{equation}\label{eq:n27}
{x_i} =  - L \cdot \cos (\pi (i - 1)/(k - 1))
\end{equation}

\begin{figure}[h!]
\begin{center}
  \includegraphics[scale = 0.9]{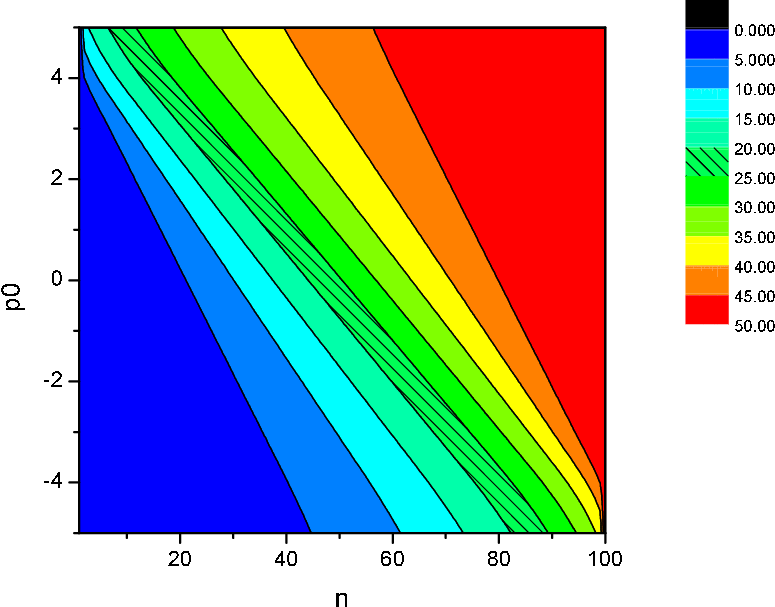} %%{pic1.pdf}
%%  \includegraphics[bb=1cm 1cm 14cm 11cm]{pic1con.eps} %%{pic1.pdf}
%%\vspace{1.cm}       % Give the correct figure height in cm
\caption{ Линии уровня для положений дислокаций относительно левого края зерна в (в микронах, $(L + x(n))/d$) заданы цветовой шкалой как функция номера дислокации и наведённого давления $p0 = pd/Db$ ($d = 1\mu $)}
\label{fig:03}       % Give a unique label
\end{center}
\end{figure}
\begin{figure}[h!]
\begin{center}
  \includegraphics[scale = 0.9]{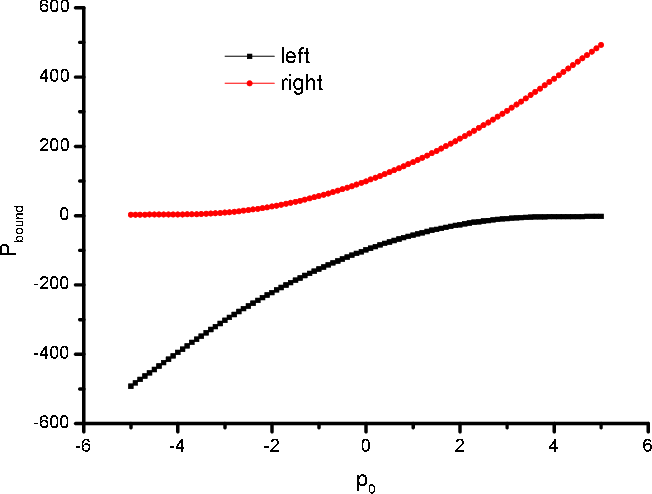} %%{pic2.pdf}
%%  \includegraphics[bb=1cm 1cm 12cm 8cm]{pic4con.eps} %%{pic2.pdf}
%%\vspace{1.cm}       % Give the correct figure height in cm
\caption{ Давление ${P_{{\rm{bound}}}}$ на  границах зерна как функция давления внутри ${p_0}$. Давление нормировано на туже величину, как на предыдущем рисунке. Коэффициент ослабления примерно в сто раз (совпадает с числом дислокаций).}
\label{fig:04}       % Give a unique label
\end{center}
\end{figure}

Эта формула показывает, как распределена система дислокаций с препятствиями на концах в отсутствие внешнего давления. Численное решение(\ref{eq:n25}) в модели дискретных дислокаций соответствует системе $\sigma ({x_i}) + p = 0$, и имеет немного другое расположение дислокаций, чем следует из формулы для непрерывного распределения(\ref{eq:n26}). Отметим, что краевые дислокации удерживаются граничными препятствиями, которые сжимают систему с обоих сторон, что компенсируется неравномерным расположением дислокаций. Все внутренние дислокации находятся в нулевом поле напряжений (если $p = 0$). Если на кристалл наложено внешнее давление, то поле напряжений внутри кристалла сдвигается на константу так, что $\sigma ({x_i})= - p$. Одновременно с этим дислокации вынуждены найти новое положение равновесия. Здесь мы опускаем вопрос о скорости релаксации в новую равновесную конфигурацию. Зависимость положений дислокаций от параметра $p$ даны на рисунке \ref{fig:03}. Давление влияет на энтропию дислокаций через их расположение.  Когда $p = 0$, распределение дислокаций симметрично, причем к краю происходит их уплотнение. Зависимость давления на границах кристалла от сдвига поля напряжений внутри на константу $p$ приведена на рисунке \ref{fig:04}. Обратим внимание, как приложенное внешнее напряжение редуцируется в равновесное напряжение внутри зерна (рис.\ref{fig:04}). Во-первых, оно постоянно по всей длине, во-вторых, поле дислокаций вынужденно компенсировать наведенное напряжение $p$, которое гораздо меньше, чем приложенное снаружи поле напряжений сдвига ${P_{{\rm{bound}}}}$. Коэффициент ослабления оказывается порядка числа дислокаций.

\begin{figure}[h!]
\begin{center}
  \includegraphics[scale = 0.9]{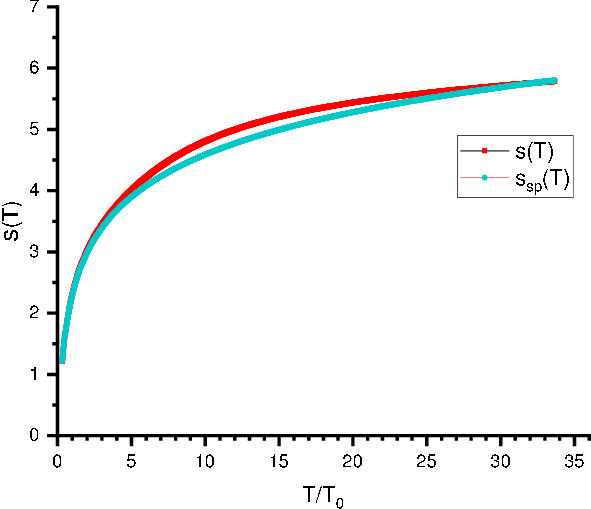} %%{doc3.pdf}
%%  \includegraphics[bb=1cm 1cm 12cm 8cm]{pic5con.eps} %%{doc3.pdf}
%%\vspace{1.cm}       % Give the correct figure height in cm
\caption{$s(T)$ - энтропия в пересчёте на одну дислокацию в зависимости от нормированной температуры, и ${s_{sp}}(T)$ - энтропия математического маятника с частотой, равной частоте колебаний дислокации на первом уровне энергии. Случай ${M_s} = 1$. Формулы для определения энтропии-(\ref{eq:n23})}
\label{fig:5}       % Give a unique label
\end{center}
\end{figure}
\begin{figure}[h!]
\begin{center}
  \includegraphics[scale = 0.9]{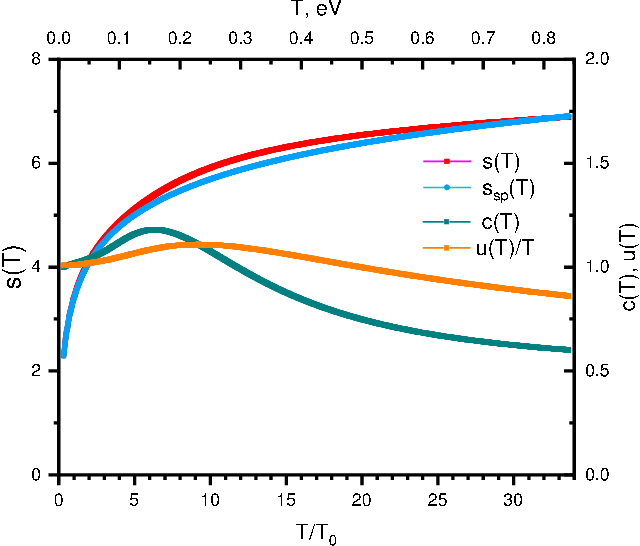} %%{doc4.pdf}
%%  \includegraphics[bb=1cm 1cm 14cm 8.5cm]{pic6con.eps} %%{doc4.pdf}
%%\vspace{1.cm}       % Give the correct figure height in cm
\caption{ $s(T)$  - энтропия в пересчёте на одну дислокацию в зависимости от температуры и ${s_{sp}}(T)$ - энтропия математического маятника с частотой, равной частоте колебаний дислокации на первом уровне энергии. Величины $c(T),u(T)/T$ - это теплоёмкость и внутренняя энергия, отнесённая к температуре, в пересчёте на одну дислокацию. Здесь взят случай ${M_s} = 9$. }
\label{fig:6}       % Give a unique label
\end{center}
\end{figure}
\begin{figure}[h]
\begin{center}
  \includegraphics[scale = 0.9]{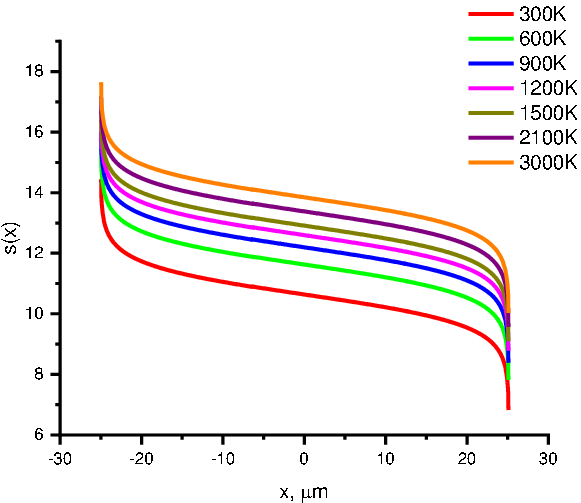} %%{doc4.pdf}
%%  \includegraphics[bb=1cm 1cm 13cm 8.5cm]{pic8con.eps} %%{doc4.pdf}
%%\vspace{1.cm}       % Give the correct figure height in cm
\caption{ $s(T)$ - энтропия в пересчёте на одну дислокацию в зависимости от координаты $x$ внутри зерна на разных изотермах. Энтропия рассчитана с учётом приложенного давления и вклада обоих слагаемых в свободную энергию (согласно(\ref{eq:nn8}) и (\ref{eq:n23})). }
\label{fig:7}       % Give a unique label
\end{center}
\end{figure}

Расчёт энтропии по формулам (\ref{eq:n22},\ref{eq:n23}) приведён на рисунке \ref{fig:5}. Данный график соответствует одномерному случаю, когда масса сердцевины дислокации взята равной ${M_s} = 1$ (в единицах массы атома титана). Проведено сравнение с моделью квантового осциллятора в параболическом потенциале. В формулах(\ref{eq:n23}) энтропия этого осциллятора, вычислена согласно\cite{t4}, и содержит индекс , означающий "simple pendulum". В данном контексте энтропия и прочие величины приведены в расчёте на одно звено вдоль сердцевины дислокации. Поскольку колебания в рассматриваемой плоскости возможны в двух измерениях, то колебательных степеней свободы на дислокацию равно 2. То есть результаты одномерных вычислений следует умножать на это число. Согласно \cite{MCh}, радиус сердцевины равен $5b$ (см.(\ref{eq:nn3})). Поэтому для оценки массы сердцевины подходит число ${M_s} = 9$. Оно примерно равно $5\sqrt \pi$. Графики энтропии для случая ${M_s} = 9$ (на одну степень свободы и в пересчёте на одну дислокацию) приведены на рисунке \ref{fig:6}. Прослеживается рост энтропии на единицу по сравнению с рис.\ref{fig:5}. На рисунке \ref{fig:6}, наряду с энтропией в модели физического маятника, приведены также результаты вычислений по модели простого маятника. Также приведены расчёты для теплоёмкости $c(T)$ и внутренней энергии физического маятника, отнесённой к температуре,$u(T)/T$  (см.(\ref{eq:n24})). Отметим, что при рассматриваемых температурах теплоёмкость простого маятника равна 1.

Как видно из расчёта у физического маятника теплоёмкость с ростом температуры проходит через максимум, а затем спадает. Асимптотика функции отлична от аналога в модели простого маятника. Также необычно ведёт себя и внутренняя энергия. Она насыщается с ростом температуры, поскольку основная часть полной энергии уходит в канал поступательного дрейфа. Качественно различия между энтропией физического и математического маятника весьма заметны, и их асимптотика существенно различна. Количественно различия существенны в диапазоне $5 < T/{T_0} < 30$ (${T_0} = 300K$) и вновь возникают при более высоких температурах, которые соответствуют нагреву порядка 1eV.  Однако, поскольку разница в рабочем диапазоне не превышает $\Delta s \le 0.3$, то для экономии времени, можно пользоваться простыми формулами из термодинамики твёрдого тела, взяв за ``средне - геометрическую частоту'' частоту колебаний сердцевины дислокации.
Расчёт локальной энтропии, с учётом вклада в свободную энергию всех слагаемых (согласно (\ref{eq:nn8}) и (\ref{eq:n23})), представлен на рисунке \ref{fig:7}. Выбран случай, когда приложено сильное внешнее поле напряжений. Внутри зерна дислокации компенсируют уменьшенное поле напряжений, равное $p = {p_{\max }} = kbD/L$ (см. (\ref{eq:n28})-(\ref{eq:n29})). Локальная энтропия на каждой изотерме зависит от координаты точки. Условие $s = const$ соответствует горизонтальной прямой, которая будет последовательно пересекать ряд изотерм.  Изоэнтропийное упорядочивание дислокаций означает рост температуры с одного края зерна к другому.

     \section{Обсуждение вопросов применимости}
 Поскольку термодинамика дислокаций опирается на довольно сильное утверждение о локальном равновесии, то следует пояснить какие поправки возникают при отклонении от этого принципа. Основные поправки возникают из-за потоков тепла. Согласно \cite{LL6}, изменение энтропии в единицу времени сводится к потоку энтропии через поверхность объема плюс два слагаемых, связанных потоком тепла через границу поверхности и диссипацией тепла в объеме ($\lambda $ -коэффициент теплопроводности) ${\mathop{\rm div}\nolimits} ( - {\bf{q}}/T) + {{\bf{q}}^2}/(\lambda {T^2})$. В работе \cite{EiTh} потоки становятся новыми дополнительными независимыми переменными, от которых зависит энтропия. На примере потоков тепла можно показать, что он будет удовлетворять простому релаксационному уравнения вида ${\bf{q}} + {\tau _1}{\bf{\dot q}} =  - \lambda \nabla T$, поэтому неравновесная энтропия оказывается ниже равновесной на величину ${s_{neq}} = {s_{eq}} - ({\tau _1}{\rho ^{ - 1}}/2\lambda T){\bf{q}} \cdot {\bf{q}}$. Это означает, что стационарная неравновесная температура ниже равновесной.  Учёт данной поправки и расчёт локально равновесной энтропии позволяет экстраполировать результаты на неравновесный случай. В модели расширенной необратимой термодинамики\cite{EiTh}, наряду с потоками тепла могут быть диссипативные потоки вязкого течения, связанные с тензором напряжений. Квадратичная форма потока девиатора тензора напряжений задаёт производство энтропии, однако не у системы дефектов, а у решётки твёрдого тела. Вязкие силы тормозят дислокации, нагревая решётку. Ещё один аргумент у функции энтропии - поток числа дислокаций. Поскольку поля напряжений между дислокациями потенциальны, то производство энтропии связано только с изменением числа дефектов, независимо от источника происхождения. В частности, для элементов объёма, где нет точек роста дислокаций, только разность потоков дислокаций через границу будут отвечать за изменение энтропии через изменение распределения плотности дислокаций в зерне. Быстрое наложение внешних напряжений делает систему дислокаций неравновесной, работа сил вызывает рост энергии (локальной температуры). Баланс роста энтропии складывается из падения конфигурационной энтропии при сгущении дислокаций и роста их колебательной энтропии при нагревании.

     \section{Заключение}
Развит подход к вычислению конфигурационной и термодинамической энтропии дислокаций. Выполненный выше анализ в рамках заданного числа дислокаций показывает, что энтропия содержит два вклада - конфигурационный и колебательный. Последний обладает температурной зависимостью через статистическую сумму для функции свободной энергии. Свободная энергия распадается на два члена, один связан с колебаниями в гармоническом потенциале решётки с глубокими ямами около положений равновесия, другой - с колебаниями дислокаций в поле напряжений от соседних дислокаций. Потенциалы отличаются по радиусу взаимодействия, и дают разный вклад в теплоёмкость и зависимость энтропии от температуры. Взаимодействие дислокаций в поле напряжения соседних дислокаций приводит к изменению вероятности для разных позиций дислокаций по решётке. Это снижает энтропию, отсчитанную от базового уровня конфигурационной энтропии. Особый результат, полученный из анализа кривых $s(T)$, связан с определением конфигурационной энтропии. Показано, что с ростом температуры происходит насыщение энтропии. Этот предел связан с невозможностью полной энтропии взаимодействующих между собой дислокаций превышать логарифм числа состояний $\ln \frac{n}{k}$ - это уровень конфигурационной энтропии, в котором все места по решётке равновероятны.

Использование модели гармонического потенциала для расчёта колебаний дислокаций позволяет учесть неэквидистантное расположение квантовых уровней с помощью квазиклассической механики. С помощью правил квантования действия удаётся рассмотреть локализованный случай и инфинитное состояние. Колебания квантового осциллятора с гармоническим потенциалом - физический маятник. Период колебаний в этом случае - функция энергии. Зависимость частоты колебаний от энергии влияет на расстояние между уровнями. Поскольку ширина ямы ограничена междуатомным расстоянием (шагом решетки $b$), то число уровней до края ямы конечно, однако они сгущаются в сторону приближения к сепаратриссе режимов колебаний.  Как следствие - изменение свойств системы, а теплоёмкость и энтропия отличаются от случая простого линейного осциллятора. Собственные колебания маятника на основе дислокаций отличны по частоте от средней частоты колебаний решётки в кристалле. В тексте работы приведены аналитические и численные результаты расчётов для термодинамической части энтропии. В теплоёмкости обнаруживается локальный максимум, связанный со сгущением уровней энергии вблизи области с фазовым переходом в режим подвижных дислокаций. Показано, как внешнее напряжение меняет локальную энтропию по длине зерна, что связано изменением равновесной плотности дислокаций. Работа внешних сил по перемещению дислокаций в положение равновесия приводит к нагреву дислокаций.

Возвращаясь к определению энтропии, предложенному в \cite{GN1,GN2}, через скорость поступательного движения дислокации $s = \ln (\frac{c}{\upsilon })$, обратим внимание, что статистический вес или число допустимых состояний системы авторы задают как минус логарифм числа сменяемых микросостояний за единицу времени за счёт движения дислокации, когда ее скорость взята в единицах скорости звука. Возникает противоречие с  \cite{Kit}, где число допустимых состояний системы задаёт конфигурационную энтропию, и зависит от плотности дислокаций. Приведём также пример с определением меры беспорядка в квантово-механической системе. В атоме водорода, энтропия обусловлена вырождением по энергии, которое связано с существованием второго закона сохранения -момента импульса ${\bf{l}}$. Полное число состояний на уровне энергии с $n \ge 1$ равно ${n^2} = {(l + 1)^2}$, и энтропия квантовой системы из-за вырождения по углу равна ${s_n} = 2\ln (1 + l)$.

Зависимость энтропии от скорости, с нашей точки зрения, более сложная и должна  определяться через дивергенцию поля скоростей дислокаций. Когда поток дислокаций через границу объёма увеличивает их плотность, происходит падение энтропии. Кроме того, большая скорость коррелирует с большей температурой, которая, в свою очередь, приводит к росту термодинамической части (колебательной) энтропии.

\end{document}